\documentclass[preprint,12pt]{elsarticle}

\usepackage{amssymb}

\begin{document}

\begin{frontmatter}

\title{Investigating particle size effects on NMR spectra of ions diffusing in porous carbons through a mesoscopic model}

\author{Anagha Sasikumar}
\author{Céline Merlet (CIRIMAT, Université Toulouse 3 Paul Sabatier, Toulouse INP, CNRS, Université de Toulouse)}


\begin{abstract}
Characterizing ion adsorption and diffusion in porous carbons is essential to understand the performance of such materials in a range of key technologies such as energy storage and capacitive deionisation. Nuclear Magnetic Resonance (NMR) spectroscopy is a powerful technique to get insights in these systems thanks to its ability to distinguish between bulk and adsorbed species and to its sensitivity to dynamic phenomena. Nevertheless, a clear interpretation of the experimental results is sometimes rendered difficult by the various factors affecting NMR spectra. A mesoscopic model to predict NMR spectra of ions diffusing in carbon particles is adapted to include dynamic exchange between the intra-particle space and the bulk electrolyte surrounding the particle. A systematic study of the particle size effect on the NMR spectra for different distributions of magnetic environments in the porous carbons is conducted. The model demonstrates the importance of considering a range of magnetic environments, instead of a single chemical shift value corresponding to adsorbed species, and of including a range of exchange rates (between in and out of the particle), instead of a single timescale, to predict realistic NMR spectra. Depending on the pore size distribution of the carbon particle and the ratio between bulk and adsorbed species, both the NMR linewidth and peak positions can be largely influenced by the particle size.
\end{abstract}

\begin{graphicalabstract}
\includegraphics[scale=0.5]{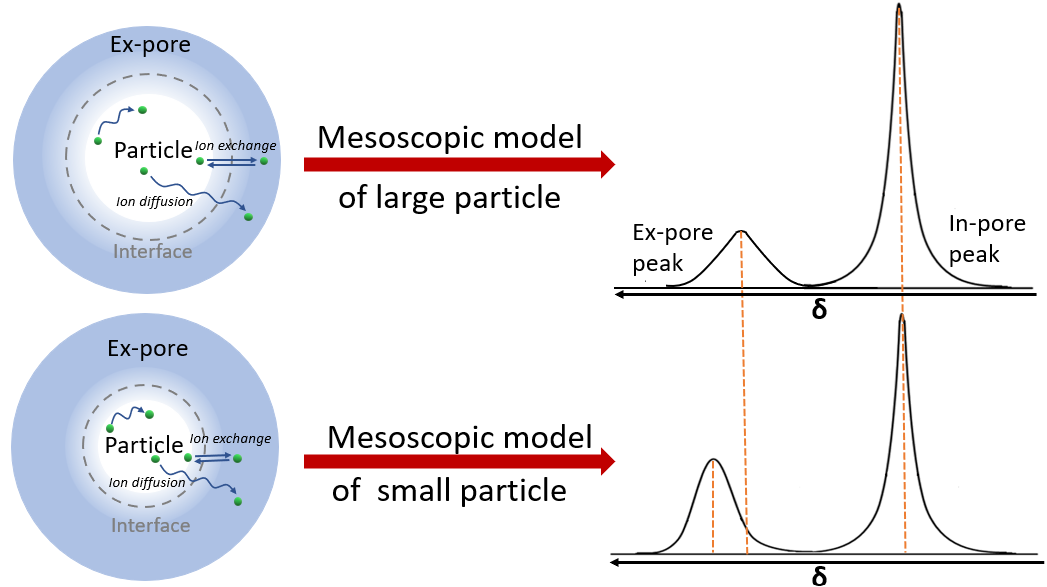}
\end{graphicalabstract}

\begin{highlights}
\item A model to predict NMR spectra of ions diffusing in particles of various sizes is developed.
\item The model allows one to consider a range of exchange rates instead of a single one.
\item The particle size can affect both the linewidth and peak positions in the NMR spectra.
\item The distribution of magnetic environments in the particle affects the linewidth.
\item The model is able to predict NMR spectra in qualitative agreement with experiments.
\end{highlights}


\end{frontmatter}

\section{Introduction}
\label{intro}

Ion transport in porous carbon materials is crucial for several key technologies including energy storage~\cite{Dou20,Luo21}, capacitive deionisation~\cite{Porada13,Luciano20}, and heterogeneous catalysis~\cite{Zhang17,Yang11}. A fundamental understanding of the adsorption and dynamic behaviour of various fluids under confinement is thus pivotal in estimating the performance of porous carbons in specific applications. 

A number of experimental techniques can be used to probe ion adsorption and transport of electrolytes at the interface with porous materials such as \emph{in situ} X-ray scattering~\cite{Prehal15,Prehal17,Prehal19}, quasi-elastic neutron scattering~\cite{Osti19} and electrochemical quartz crystal microbalance~\cite{Tsai14,Chai20,Levi10}. Amongst these experimental methods, Nuclear Magnetic Resonance (NMR) spectroscopy stands out as a very useful and non-invasive technique to study the dynamics of confined species due to its nucleus sensitivity. The applicability of NMR to study ion adsorption and transport comes from the fact that a molecule in the bulk or confined in pores of various sizes will experience different magnetic environments. The NMR spectra of electrolyte species will usually show at least two peaks corresponding to ``in-pore"  and ``ex-pore" (bulk) species~\cite{Forse13,Borchardt13}.

The existence of a range of magnetic environments in the porous carbons generally results in various linewidths for the in-pore peak depending on the similarity between environments, often related to the pore size distribution, and the motion dynamics between these different environments~\cite{levitt13,Merlet15}. As a consequence, the variation of the linewidth with electrolyte nature and temperature can be used as a qualitative probe of the adsorbed species dynamics. This was for example demonstrated in the experiments conducted by Forse~\textit{et~al.}~\cite{Forse15} in which the authors observed an increase in the linewidth with a decrease in temperature, associated with the decreased mobility of the ions. However, predicting the linewidths is difficult as a broad range of local magnetic environments and diffusion coefficients is expected owing to the heterogeneity of the pore space~\cite{Furtado11}. The diversity of pore sizes as well as their spatial connectivity~\cite{Merlet15} can also affect the transport properties, adding to the complexity of the system. 

When confined species undergo a relatively fast motion compared to the range of frequencies explored (in the various magnetic environments), the motional averaging leads to a well-defined average chemical shift~\cite{Xing14}. In such cases, it is possible to use the position of the in-pore peak to gain knowledge about the structure and porosity of the porous carbon material~\cite{Forse15b}. Indeed, determinations of Nucleus Independent Chemical Shifts (NICS) from Density Functional Theory (DFT) calculations have shown that larger chemical shifts are expected for smaller pores and for larger graphitic domains~\cite{Borchardt13,Cervini19,Forse14}. 

More specific NMR experiments, such as pulsed-field gradient (PFG) and two-dimensional exchange experiments (EXSY) can be performed to evaluate diffusion coefficients and exchange rates in a quantitative fashion~\cite{Alam16,Cui22}. Earlier PGF-NMR studies have indicated a reduced mobility of species confined in various porous carbons proportional to a reduction in the average pore size~\cite{Heink93,Dubinin88,dagostino14,Forse17}. The diffusion process is also known to be affected by the tortuosity of the porous material~\cite{odintsov98} and the variation in ion concentration, for example when applying an electric potential to the carbon~\cite{Forse17}.

2D EXSY experiments are very useful to provide additional information on the exchange of species between distinct environments. An in-pore/ex-pore exchange in samples saturated with organic electrolytes and commercially available YP-50F carbon was observed at a millisecond timescale in independent EXSY experiments conducted by Griffin~\textit{et~al.}~\cite{Griffin14} and Fulik~\textit{et~al.}~\cite{Fulik18}. It was proposed that the in-pore/ex-pore exchange process consisted of a ``slow" component attributed to the exchange between species present in the ``deep" interior region of the carbon and the bulk, and a ``fast" component attributed to the exchange between the species present at the ``surface" region of the carbon, close to the bulk, and the bulk~\cite{Fulik18}. A later study~\cite{Fuentes21} indicated inter-pore exchange at a mixing time of 5~ms and an in-pore/ex-pore exchange at a mixing time of 50~ms for aqueous LiCl electrolyte in mesoporous carbons underlining the difference in exchange rates for processes occurring at different levels.

Multiple dynamical processes characterised by different timescales imply an influence of the particle size on the NMR spectra. Studies conducted on particles with sizes of a few micrometers indicated a chemical shift averaging and a consequent change in the linewidth~\cite{Forse13} which was not observed in experiments that used large particles~\cite{Cervini19}. However, the effect of particle size on the NMR spectrum has rarely been systematically investigated. Cervini~\textit{et~al.}~\cite{Cervini20} conducted experiments on particles with two different sizes which revealed the presence of an exchange peak associated with the ex-pore environment. In addition, smaller rate constants were estimated for the slow and fast in-pore/ex-pore exchange processes for large particles compared to the small particles which were reflected in the broader peak for the smaller particle size. The observations were associated with the shorter diffusion paths from the interior to the exterior part for the smaller particles leading to faster exchange.   

Clearly, the ion mobility in porous carbons is complex, involving diverse processes that can affect the NMR spectra. Common analytical models~\cite{Cavanagh,Keeler} describing exchanges between two sites with well-defined chemical shifts are not sufficient to explain the shape and peak positions of experimental NMR spectra. A mesoscopic model has been proposed earlier in our group to predict NMR spectra of species diffusing in porous carbons~\cite{Merlet15}. The model incorporates microscopic information from molecular dynamics simulations and DFT calculations to evaluate the NMR signal. It was shown previously that the model could predict satisfactorily the average chemical shifts of several electrolyte species in contact with mesoporous and nanoporous carbons~\cite{Merlet15,Sasikumar21}. An advantage of this method is that various factors affecting the spectra can be considered simultaneously or independently. However, the model was previously not adapted to include in-pore/ex-pore exchange, being limited to a description of the in-pore environments.  In the present study, we improve the model by including a bulk region and apply it to perform a systematic assessment of the effect of particle size on the NMR spectra.

\section{Methods}

\subsection{Description of the mesoscopic model}

We adapt and use a lattice-based model to explore the particle size effects on the NMR spectra of ions diffusing through, and in and out of porous carbons. The carbon particle model, described in earlier works~\cite{Merlet15,Anouar19,Sasikumar21,Sasikumar22}, uses a coarse-grained approach in which the carbon particle is represented as a collection of slit pores accessible to the electrolyte (or more generally fluid) species. The properties of the pore corresponding to each lattice site (i.e., the width of the pore and the surface area of the lateral walls) need to be specified. The quantities of ions in the different pores is determined using adsorption energies extracted from molecular dynamics simulations. The resonant frequencies of the considered specie at a given lattice site, i.e. in a given pore of well defined size, is determined using NICS data from DFT calculations. 

The diffusion of electrolyte species across lattice sites is governed by an acceptance rule by which a move from site~\textit{i} to site~\textit{j} is accepted with a probability:
\begin{equation}\label{probability}
P(i\to j) = \left\{
\begin{array}{cc}
\exp\left ( \frac{-(E_j-E_i)}{{\rm k_B}T} \right ) & \mathrm{if\ } E_j>E_i \\
1 & \mathrm{if\ } E_j\leq {E_i}
\end{array}
\right.
\end{equation}
where $E_i$ is the energy assigned to site $i$ (corresponding to a quantity of ions), ${\rm k_B}$ is the Boltzmann constant and $T$ is temperature of the system. To describe the possible existence of bottlenecks between pores or slower diffusion in some circumstances, an additional energy barrier, which will be designated as an activation energy $E_a$ in the remainder of this article, can be added in the model. In this case, the probability of a move from \textit{i} to \textit{j} to happen is reduced by $\alpha_{ij}=\exp(-E_a/{\rm k_B}T)$, leading to a similar decrease in diffusion.

Such a model only considers the exchange of ions between pores inside the particle (in-pore exchange) but not the exchange of ions between the surrounding bulk electrolyte and the pores of the carbon (ex-pore~/ in-pore exchange). In the new set-up, the lattice sites are assigned as ``bulk" or ``particle" depending on their position in the particle-liquid system. All sites within the specified particle region, defined as a spherical particle (size given in lattice units), are considered as belonging to the carbon particle. All other sites are regarded as bulk electrolyte. Due to the design of the model, there is still a volume, corresponding to a parallelepiped with a given surface and width, associated with each site in the bulk region. It is worth noting that both the lattice and the particle sizes have to be chosen large enough that i)~the system has a sufficient amount of bulk region to capture the in-pore~/ ex-pore exchange and ii)~the pore size distribution of the carbon is represented satisfactorily. 

\begin{figure}[ht!]
\centering
\includegraphics[scale=0.6]{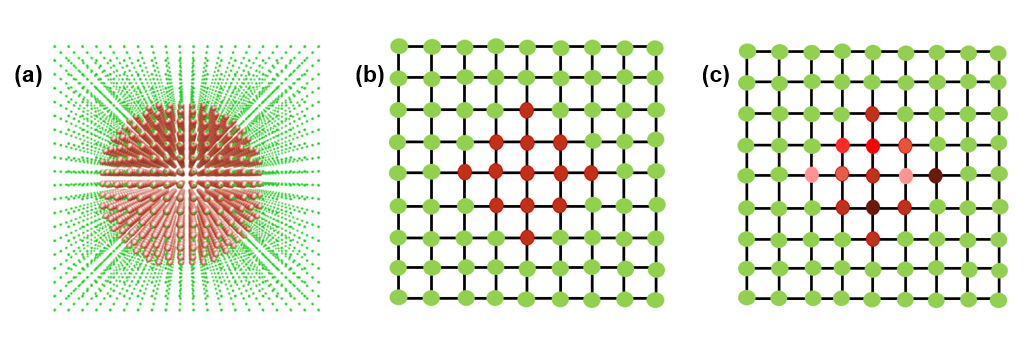}
\caption{Schemes illustrating (a) the lattice representation of the carbon particle (in red) and the electrolyte (in green) for a 20$\times$20$\times$20 lattice size, (b) the implementation of the 2-site model with homogeneous values for the chemical shifts in sites corresponding to the bulk electrolyte (green) and the carbon particle (red) and, (c) the implementation of a realistic carbon particle model with a single value for the chemical shifts in sites corresponding to the bulk electrolyte (green) and various chemical shifts in sites corresponding to the carbon particle (various shades of red).}
\label{particle-model}
\end{figure}

The assignment of the parameters corresponding to each site is performed in two ways represented in Figure~\ref{particle-model}.
\begin{itemize}
\item 2-site model: In this model, all the pores (and parallelepiped for the bulk sites) in the lattice are identical with a given pore surface and pore width. As a consequence, all the quantities of adsorbed ions (in relation with the site energies) are equal. For bulk sites, the chemical shift is set to 0~ppm. For particle sites, the chemical shift is set to a non-zero constant.
\item Realistic model: In this model, the pore widths for the particle sites are set following a realistic pore size distribution corresponding to the carbon material under consideration. For bulk sites, the pore width corresponding to the largest pores in the pore size distribution is considered. The site surfaces are set following a log-normal function with a mean of -0.1 and a standard deviation of 0.25 as described in earlier works~\cite{Merlet15,Anouar19,Sasikumar21}. For bulk sites, the chemical shifts are set to 0~ppm. For particle sites, chemical shifts are assigned according to the pore widths.
\end{itemize}

\begin{figure}[ht!]
\centering
\includegraphics[scale=0.6]{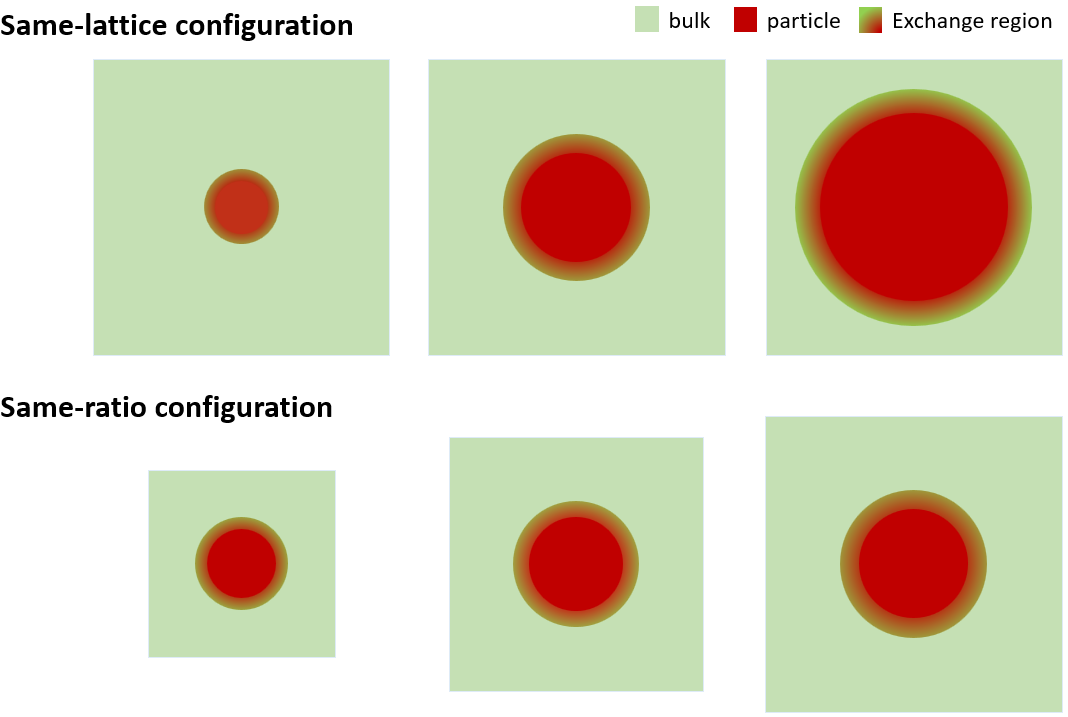}
\caption{Schemes illustrating the lattice models corresponding to the same-lattice configuration (top) and same-ratio configuration (bottom) for different particle sizes. The green area represents the bulk electrolyte, the red area represents the particle and the gradient from red to green represents the in-pore~/~ex-pore exchange region.}
\label{def-same-lattice-ratio}
\end{figure}

In both models, simulations can be performed for two types of configurations represented in Figure~\ref{def-same-lattice-ratio}.
\begin{itemize}
\item Same-lattice: In this configuration, the size of the lattice is fixed while the particle size is varied. In this case, the carbon/electrolyte system is represented using a lattice of size 20$\times$20$\times$20. Particle sizes ranging from 6 to 16 lattice units are investigated. 
\item Same-ratio: In this configuration, both the lattice and particle sizes are varied. Here, the ratio between the number of sites representing the carbon particle and total number of sites is set to approximately 0.5. For example, for a particle size of 6, a lattice of size 12$\times$12$\times$12 is considered while for a particle size 16, a lattice of 32$\times$32$\times$32 is used.
\end{itemize}

For all simulations, 50,000~steps are performed  with a time\-step of 5~\textmu s. As in previous works~\cite{Merlet15,Anouar19,Sasikumar21}, these parameters are sufficient to observe a complete decay of the NMR signal. The spectra are generated and the chemical shifts and linewidths are assessed. 

\subsection{Parametrisation of the mesoscopic 2-site model}

In this study, we focus on calculating the $^{19}$F NMR spectra of BF$_4^-$ anions in an electrolyte composed of the 1-butyl-3-methylimidazolium tetrafluoroborate salt dissolved in acetonitrile (1.5~M [BMI][BF$_4$] in ACN) in contact with a model carbon particle. For the 2-site model, the particle is loosely inspired from the one of a porous carbon generated \textit{in silico} by Deringer~\textit{et~al.}~\cite{Deringer18}, namely  GAP930 which has a structure relatively close to carbide derived carbons synthesised experimentally.

The pore (or bulk sites) widths are set to 35~nm, the largest pore size identified in the pore size distribution of GAP930 carbon obtained with Poreblazer~\cite{poreblazer}. The pore surfaces are set to 0.981~nm$^2$, the surface area of the circumcoronene molecule, following previous works~\cite{Merlet15,Xing14}. According to these pore parameters, a constant free energy of 33.75~kJ~mol$^{-1}$ is set for all sites. Details for the definition of energies and quantities of ions can be read in previous works~\cite{Merlet15,Anouar19,Sasikumar21,Sasikumar22}. It is worth noting that in this case, all energies being equal, the actual value chosen will not affect the results as it is energy differences and not absolute values which influence the site populations and dynamics.

A value of -7.2~ppm is chosen for the resonant frequencies at particle sites. This is the average chemical shift obtained for in-pore anions in a [BMI][BF$_4$]/ACN - GAP930 system, evaluated by performing a classical full-particle simulation~\cite{Merlet15,Anouar19,Sasikumar21,Sasikumar22} where all sites have distinct frequencies.

\subsection{Description of the analytical model and comparison with the 2-site mesoscopic model}

A simple analytical model exists for the prediction of NMR spectra for an exchange of species between two chemically dissimilar environments with frequencies $\omega_1$ and $\omega_2$. When the exchange is in the fast regime, that is when the difference in resonant frequencies, $\Delta\omega~=~|\omega_1~-~\omega_2|$, is lower than the exchange rate between the sites, $k_{exc}$, the NMR signal is given by~\cite{Cavanagh}:
\begin{equation}
\label{FT-fast}
M(t) = \exp[(-i\overline{\Omega}+\overline{\rho}+p_1p_2~\Delta\omega^2/k_{exc})t]
\end{equation}
where $\overline{\Omega}$ is the average resonance frequency, $\overline{\rho}$ is the average relaxation decay constant, and $p_1$ and $p_2$ are the fractions of species in the two sites.  

In order to compare the spectra obtained with the 2-site mesoscopic model proposed here and the results from the analytical model, it is necessary to estimate exchange rates between in-pore (particle) and ex-pore (bulk) regions. An exchange between the in-pore and bulk environment occurs when the anion moves from a particle site to a bulk site or vice versa. The overall exchange rate can be evaluated by determining the fraction of all moves resulting in a change in environment (chemical shift) at a given timestep:
\begin{equation}
\label{exchange-rate}
k_{exc} = \frac{n_{exc}}{n_{tot}}\times\frac{1}{\tau}
\end{equation}
where $n_{exc}$ is the number of moves resulting in an in-pore~/~ex-pore exchange at each timestep, $n_{tot}$ is the total number of moves at each timestep and $\tau$ is the characteristic timestep of the lattice. In the mesoscopic model, there are no additional decay processes such that, for comparison purposes, the relaxation decay constant is taken to be 0~s$^{-1}$.

\subsection{Parametrisation of the mesoscopic realistic model}

For a more realistic representation of the [BMI][BF$_4$]/ACN - GAP930 system, pore parameters for all sites representing the particle are set to be distinct. The pore size distribution of GAP930 determined using Poreblazer~\cite{poreblazer} (see Figure~\ref{psd}) is used to determine the widths of the pores representing the particle. The widths for the sites representing the bulk are set to 35~nm. The pore surfaces follow a log-normal distribution as described previously. The site energies (and corresponding quantities of ions) and chemical shifts are determined according to the pore widths and surfaces as in previous works~\cite{Merlet15,Anouar19,Sasikumar21}. The distribution of chemical shifts obtained for a particle of size 10 and a 20$\times$20$\times$20 lattice is given in Figure~\ref{psd}.

\section{Results and discussion}

\subsection{Investigation of particle size effects using a 2-site representation}

\begin{figure}[ht!]
\centering
\includegraphics[scale=0.24]{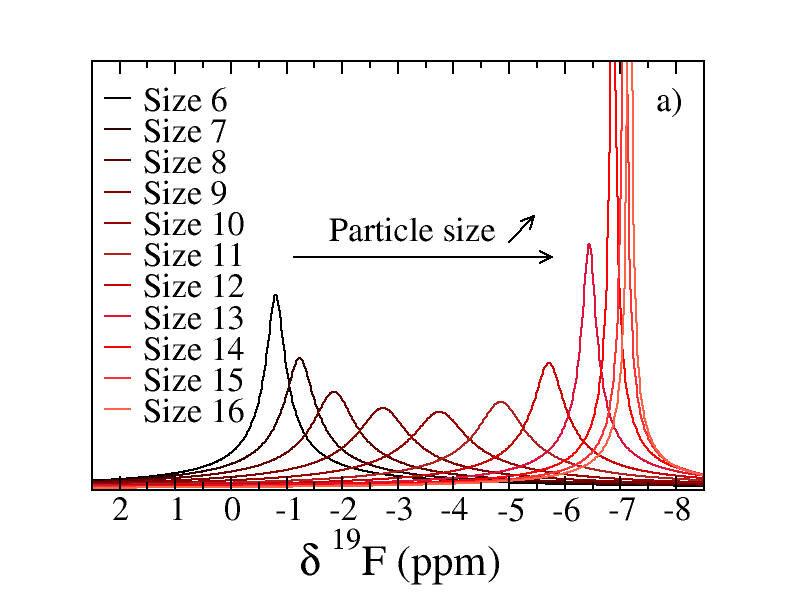}
\includegraphics[scale=0.24]{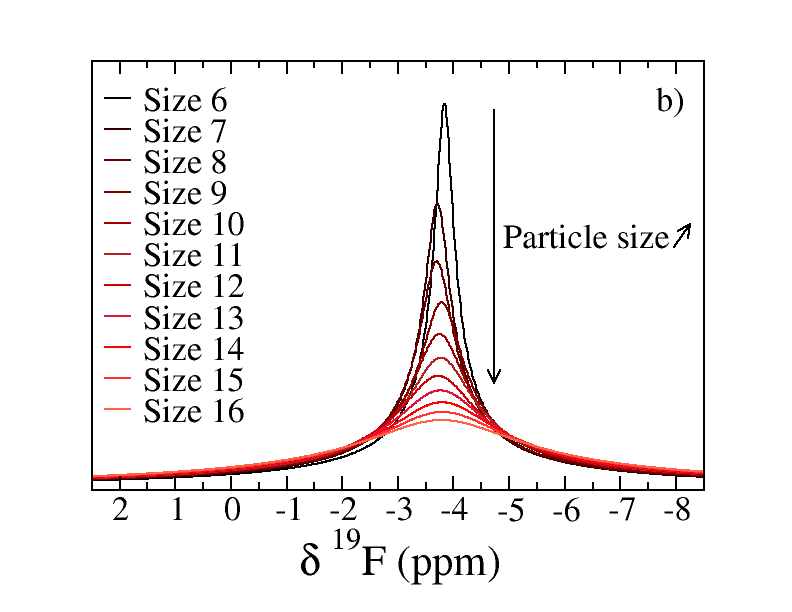}
\caption{a) $^{19}$F NMR spectra of BF$_4^-$ anions in the [BMI][BF$_4$]/ACN system, in the ``same-lattice" 2-site configuration, for various particle sizes (in lattice units). The chemical shifts of the in-pore (respectively bulk) species are set to a constant value of $-7.2$~ppm (respectively 0~ppm). b) Same as a) but for the ``same-ratio" 2-site configuration.}
\label{NMR-spectra-2-site}
\end{figure}

As a first way to understand the effect of particle size on the NMR spectra of species diffusing in and out of carbon particles, we focus on a 2-site model where the chemical shifts of the in-pore species are set to -7.2~ppm while the bulk species have a chemical shift equal to 0~ppm. The particle size is varied and the NMR spectra are determined. Within that representation, the ``same-lattice" and ``same-ratio" setups are explored. The predicted NMR spectra for the ``same-lattice" and ``same-ratio" systems for particle sizes of 6 to 16 lattice units are shown in Figure~\ref{NMR-spectra-2-site}. 
With the parameters chosen, all the NMR spectra show a single peak at variable positions and with different linewidths.

\begin{figure}[ht!]
\centering
\includegraphics[scale=0.24]{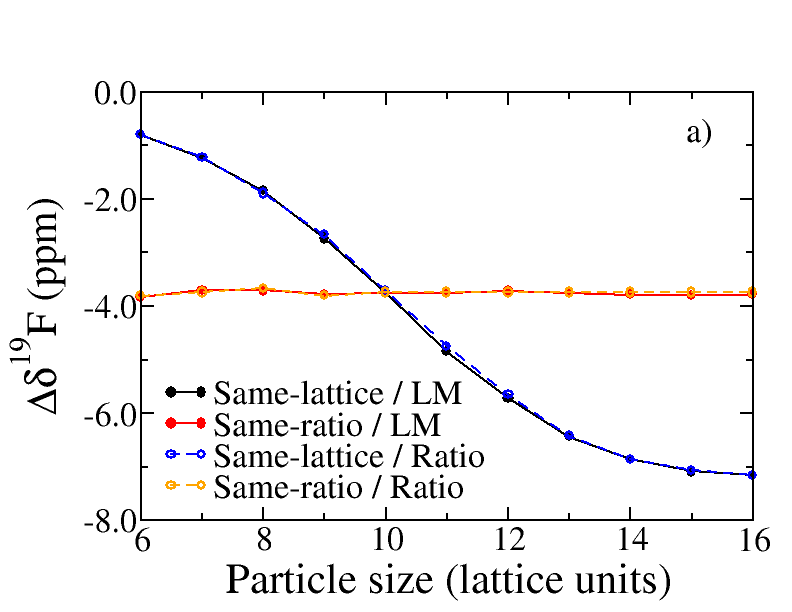}
\includegraphics[scale=0.24]{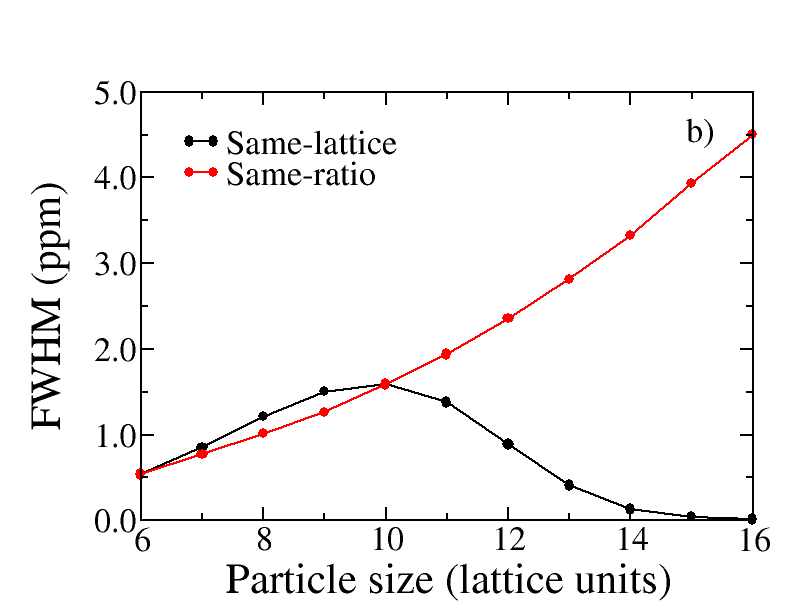}
\caption{Peak positions (a) and full widths at half maximum (b) for $^{19}$F NMR spectra of BF$_4^-$ anions in the [BMI][BF$_4$]/ACN 2-site systems. The peak positions determined using the lattice model are in agreement with simple estimations from the ratios between in-pore and bulk sites.}
\label{Dd-fwhm-2-site}
\end{figure}

The fraction of in-pore species and the effective exchange rates between in-pore and bulk sites are given in Table~\ref{in-out-ratios-rates}. It is worth noting that the fraction of in-pore sites is not exactly 0.5 for cases where the particle diameter is half of the lattice size, this is due to the spherical shape of the particle in a discrete lattice. As expected the peak position, determined by the ratio between in-pore and ex-pore species, is almost constant for the ``same-ratio" representation while it varies significantly for the ``same-lattice" system. This is clearly shown in Figure~\ref{Dd-fwhm-2-site}. Unsurprisingly, the peak positions determined through the lattice model are in agreement with simple estimations based on the ratio between in-pore and bulk species.

\begin{table}[h]
\begin{center}
\begin{tabular}{|c|c|c|c|c|}
\hline 
 & \multicolumn{2}{|c|}{Same-lattice} & \multicolumn{2}{|c|}{Same-ratio} \\
\hline 
Particle size & fraction of & $k_{exc}$ & fraction of & $k_{exc}$ \\
(lattice units) & in-pore species & (kHz) & in-pore species & (kHz) \\
\hline
6 & 0.11 & 57 & 0.53 & 255\\
\hline
7 & 0.17 & 75 & 0.52 & 213\\
\hline
8 & 0.26 & 99 & 0.51 & 190\\
\hline
9 & 0.37 & 127 & 0.53 & 172\\
\hline
10 & 0.52 & 157 & 0.52 & 157\\
\hline
11 & 0.66 & 144 & 0.52 & 141\\
\hline
12 & 0.79 & 120 & 0.52 & 127\\
\hline
13 & 0.89 & 89 & 0.52 & 120\\
\hline
14 & 0.95 & 56 & 0.52 & 111\\
\hline
15 & 0.98 & 21 & 0.52 & 105\\
\hline
16 & 0.99 & 7 & 0.52 & 97\\
\hline
\end{tabular}
\end{center}
\caption{Fractions of in-pore sites and exchange rates for different particle sizes in the ``same-lattice" and ``same-ratio" configurations (2-site model).}
\label{in-out-ratios-rates}
\end{table}

Figure~\ref{Dd-fwhm-2-site} also shows the full widths at half maximum (FWHM) of the NMR spectra calculated. The trends are very different for the ``same-lattice" and ``same-ratio" systems. This is not surprising as the linewidth is expected to depend both on the fractions of in-pore and bulk sites as well as the exchange rates between these two types of sites. In the same-ratio case, the ratio is almost constant but the larger the particle, the slower the exchange rate (see Table~\ref{in-out-ratios-rates}). This is due to the fact that the larger the particle and the lattice, the larger the volumes of in-pore and bulk regions. As a consequence, the residence time of ions in each of these regions increases when the particle size increases. In that case, the linewidth increases with the particle size as the exchange rate brings the system closer to the coalescence frequency.

\begin{figure}[ht!]
\centering
\includegraphics[scale=0.24]{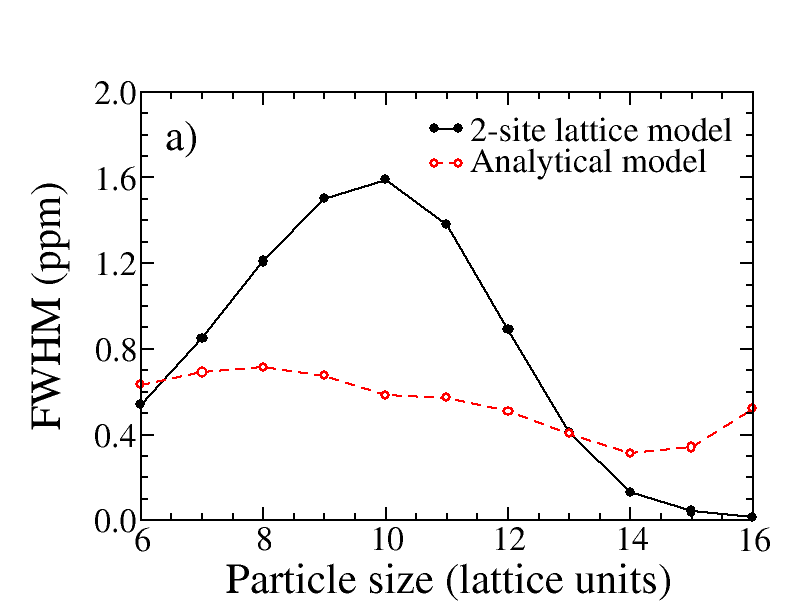}
\includegraphics[scale=0.24]{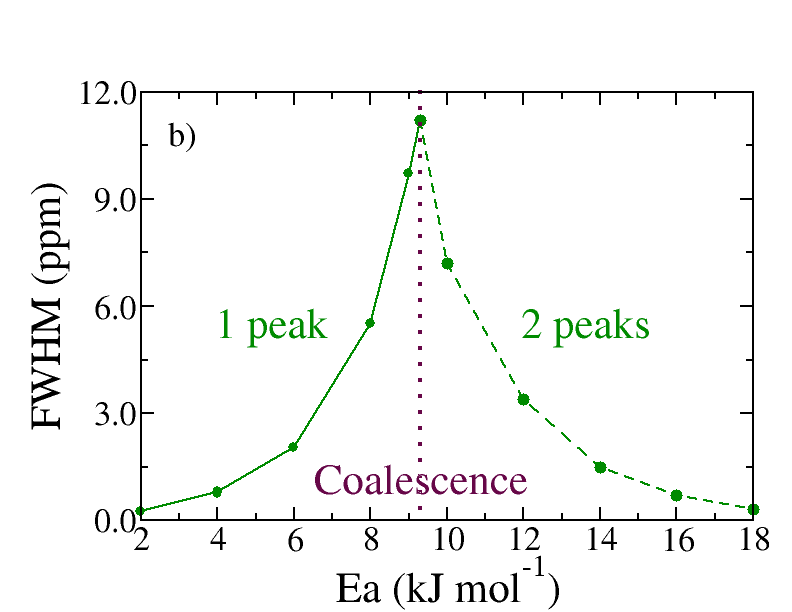}
\caption{a) Full widths at half maximum for the 2-site ``same-lattice" lattice model and for a comparative analytical model. b) Determination of the activation energy corresponding to the coalescence point for a model equivalent to analytical approaches.}
\label{fwhm-2-site-analytical}
\end{figure}

For the ``same-lattice" system, the FWHM as a function of particle size shows a maximum for a particle size of 10. While this might seem intuitive as it corresponds to the case where the fractions of in-pore and bulk sites are almost equal, the exchange rates are also varying largely when the particle size increases. To see if the linewidth could be obtained with a simple analytical model, we plot the linewidth as $8\pi p_1 p_2 \Delta\omega^2 / k_{exc}$ as proposed by Ramey~\emph{et al.}~\cite{Ramey71} and in agreement with the analytical model presented in the Methods section. Results are shown in Figure~\ref{fwhm-2-site-analytical}.

While the linewidths obtained through the analytical model are in the same range as those predicted by the lattice model, the actual values and trend with particle size are noticeably different. In fact, this is not very surprising as the analytical model only describes what happens in the case of a single average exchange rate while the lattice model represents a variety of exchange rates. It is worth noting that the exchange rates given in Table~\ref{in-out-ratios-rates} also only correspond to instantaneous exchange rates happening in the exchange region rather than effective average exchange rates across the simulation.

\begin{figure}[ht!]
\centering
\includegraphics[scale=0.6]{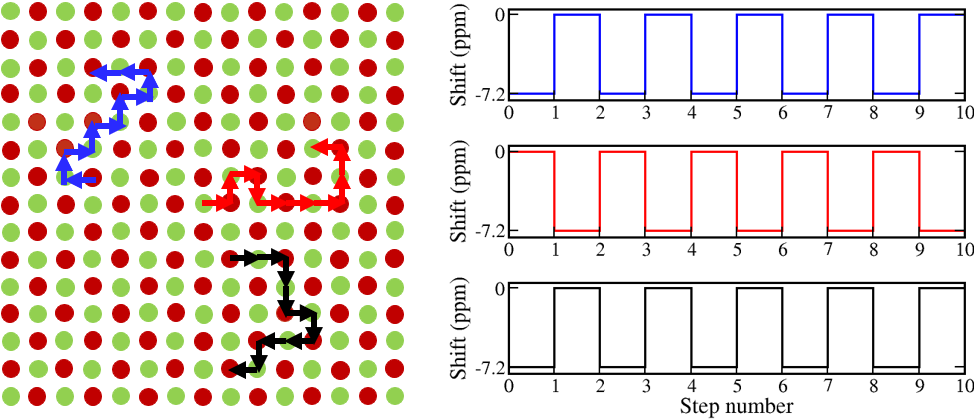}
\includegraphics[scale=0.6]{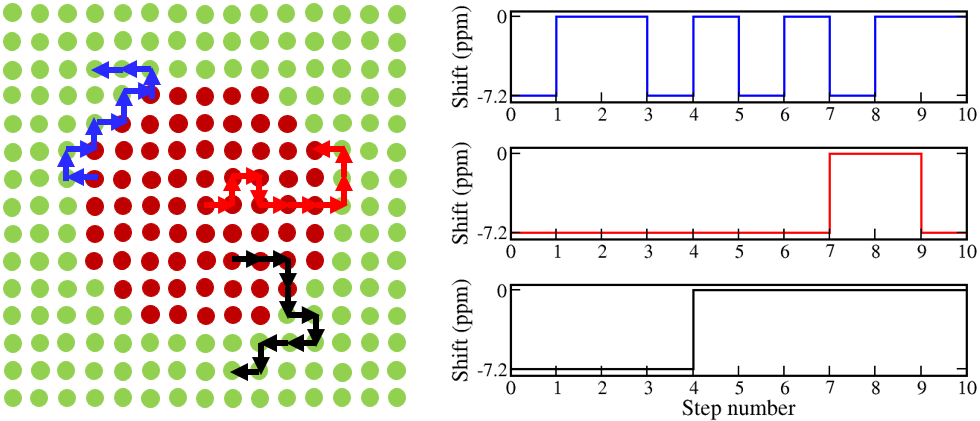}
\caption{Figure to illustrate the distribution of exchange rates corresponding to different ion diffusion paths in the case of an analytical model (top panel) and the particle lattice model (bottom panel). Ex-pore sites are in green while in-pore sites are in red. In the case of the analytical model, the ions change environment at each step and there is a single exchange rate. In the case of the lattice model, the ions can experience a wide range of diffusion paths leading to a distribution of exchange rates.}
\label{distrib-exchange-rates}
\end{figure}

Figure~\ref{distrib-exchange-rates} illustrates this disparity between the lattice model and the ana\-ly\-ti\-cal approach.
In the case of the analytical model, the ions change environment at each step and there is a single exchange rate. In the case of the lattice model, the ions can experience a wide range of diffusion paths leading to a distribution of exchange rates. Taking into account these considerations, it is not surprising that these two models are giving very different results in terms of linewidths. 

It is possible to check that the lattice model does behave as expected with respect to linewidth prediction by conducting simulations with a distribution of sites corresponding to the analytical model. In the representation chosen, corresponding to the top panel of Figure~\ref{distrib-exchange-rates}, the fraction of in-pore and ex-pore sites are equal, each corresponding to half of the lattice. By varying the activation energy, $E_a$, it is possible to tune the exchange rate and find the coalescence point. The results of these simulations for $\Delta\omega$~=~7.2~ppm are shown in Figure~\ref{fwhm-2-site-analytical}. 

Only one peak is observed for activation energies below approximately 9.3-9.4~kJ~mol$^{-1}$ (solid line) while two peaks of equal widths are seen for activation energies above this value (dashed line). Each value of activation energy corresponds to a given exchange rate:
\begin{equation}
k_{exc} = \frac{1}{\tau} \times \exp \left ( -\frac{E_a}{{\rm k_B}T} \right ).
\end{equation}
For an activation energy of 9.4~kJ~mol$^{-1}$, a temperature of 298~K and $\tau$ equal to 5~\textmu s, $k_{exc}$ is equal to 4.5~kHz which is the coalescence rate expected from:
\begin{equation}
k_{coal} = \frac{2\pi\Delta\omega}{2\sqrt{2}}
\end{equation}
in the case of $^{19}$F NMR in the considered system. This confirms that the lattice model predicts the linewidth correctly for simple cases, accessible through an analytical model, but it also allows for the study of much more complex systems.

\subsection{Application of the mesoscopic model on a realistic carbon particle}

The application of the lattice model on a 2-site example is useful to highlight the effect of particle size in idealised systems. Nevertheless, in real carbon particles, there is a distribution of pore sizes, and consequently a variety of magnetic environments with a range of exchanges rates between them. This part focuses on the application of the lattice model to more realistic systems using a specific disordered carbon, GAP930~\cite{Deringer18}, generated \textit{in silico} but with characteristics close to some experimental carbons. 

\begin{figure}[ht!]
\centering
\includegraphics[scale=0.24]{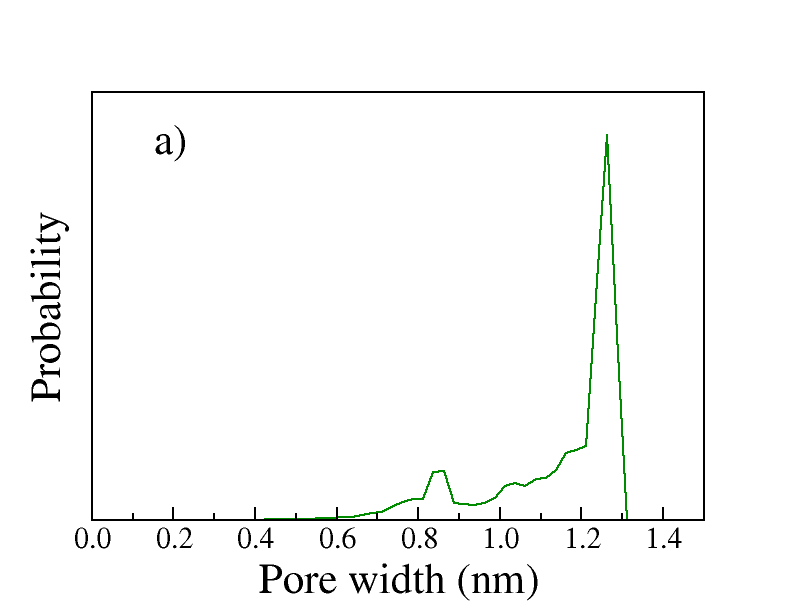}
\includegraphics[scale=0.24]{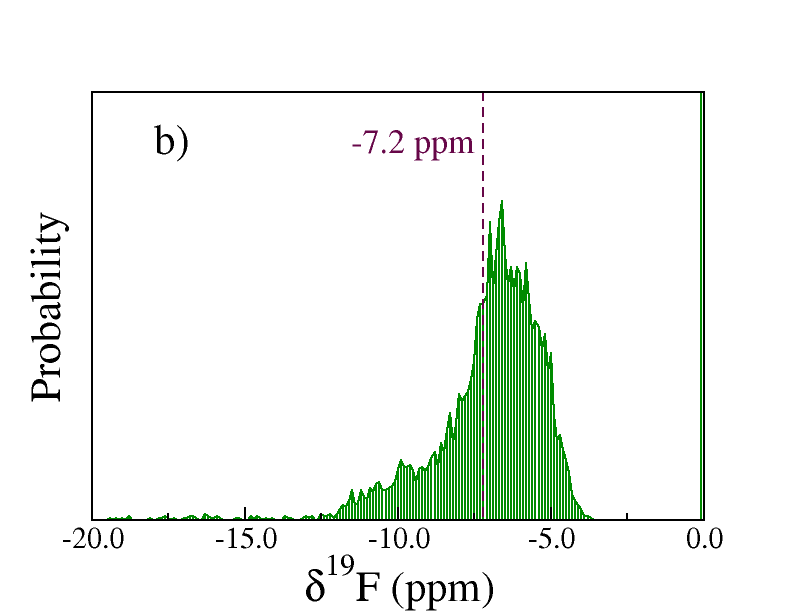}
\caption{a) Pore size distribution considered for the particles in the realistic model (corresponding to the GAP930 carbon generated by Deringer~\emph{et~al.}~\cite{Deringer18}). b) Example of chemical shift distribution for a particle of size 10 (the -7.2~ppm shift used in the 2-site model is indicated for reference).}
\label{psd}
\end{figure}

\begin{figure}[ht!]
\centering
\includegraphics[scale=0.24]{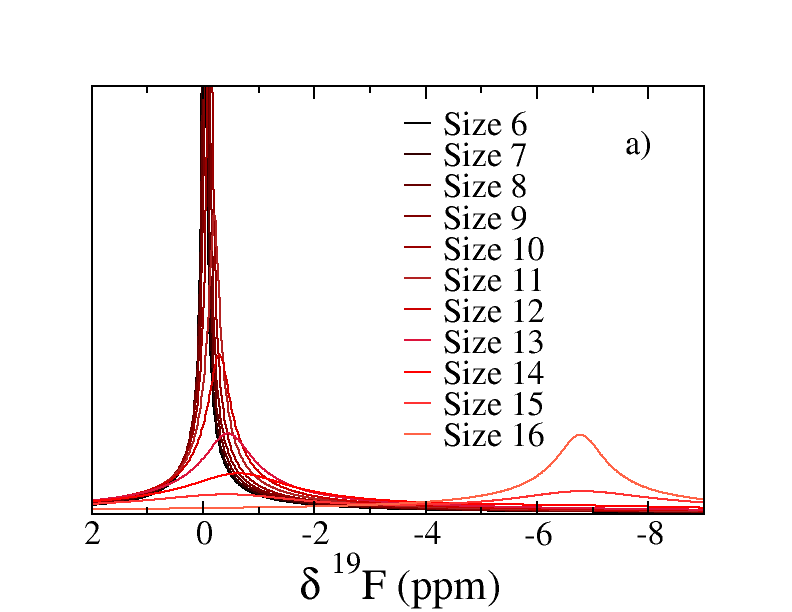}
\includegraphics[scale=0.24]{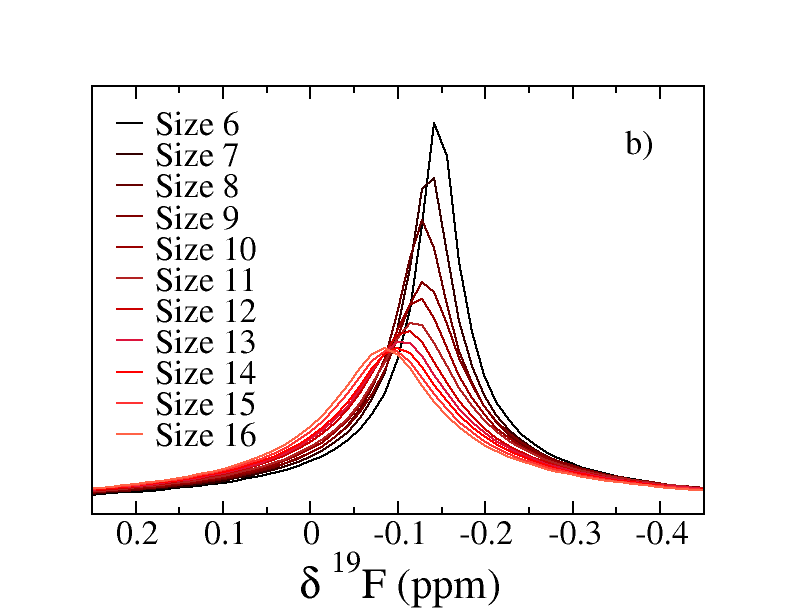}
\caption{$^{19}$F NMR spectra of BF$_4^-$ anions in the BMI-BF$_4$/ACN and GAP930 carbon system for various particle sizes in a) the ``same-lattice" configuration and b) the ``same-ratio" configuration. Note the difference in the x-axis scale for the two plots.}
\label{spectra-realistic}
\end{figure}

As described in the Methods section, the pore size distribution of the carbon particle is taken identical to the one of GAP930 calculated through Poreblazer~\cite{poreblazer}. The pore size distribution and resulting distribution of chemical shifts across the particle are shown in Figure~\ref{psd}. The NMR spectra were simulated for the ``same-lattice" and ``same-ratio" representations and results are shown in Figure~\ref{spectra-realistic}.

For the ``same-ratio" configuration, there is only one peak for all particle sizes studied and there is only a small drift of the peak position with the particle size. It is worth noting that the peak position is very different for the realistic and 2-site representations, respectively close to -0.1~ppm and -3.8~ppm. Importantly, while the ratio between particle sites and bulk sites is close to 1, the quantities are not the same in the realistic model. Indeed, in the realistic representation, there are less ions in small pores than in large bulk sites. As a consequence, it is not surprising that the peak is much closer to zero. 

The linewidth is also much smaller for the realistic model, approximately 0.1-0.2~ppm, compared to the 2-site model, approximately 1-6~ppm. The variation of full width at half maximum with particle size is much less dramatic in the realistic representation, divided by around 2 between sizes 16 and 6, compared to the 2-site model, in which it is divided by around 10. Overall, while the trends are the same between the two representations, the results show that the 2-site model is too crude a representation to get a good idea of the variation of NMR spectra with the particle size.

\begin{figure}[ht!]
\centering
\includegraphics[scale=0.24]{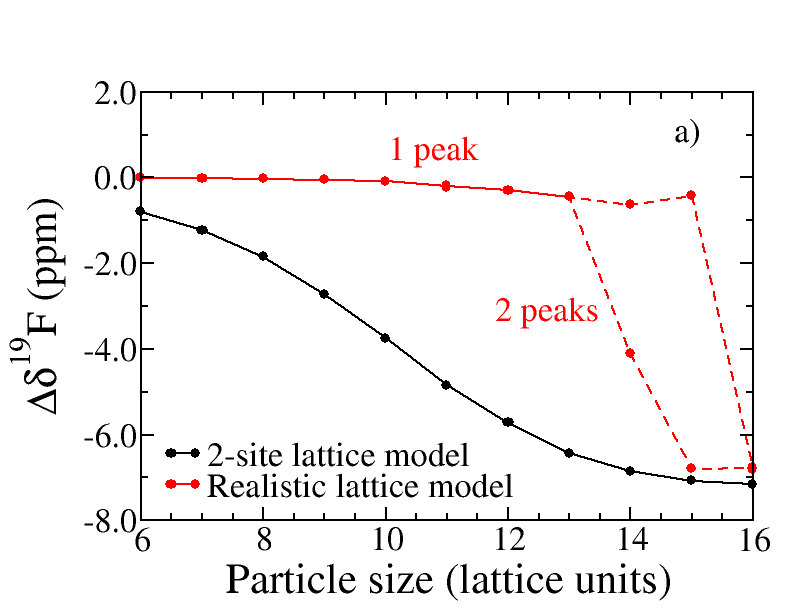}
\includegraphics[scale=0.24]{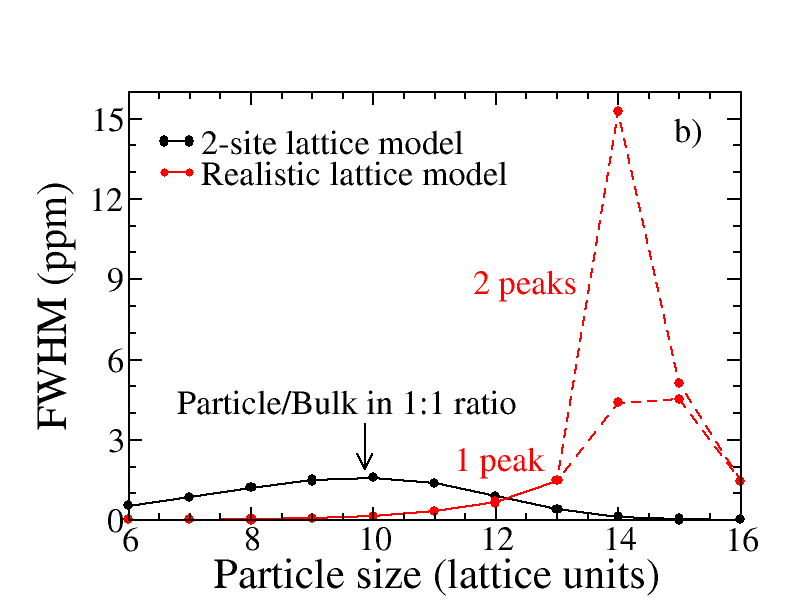}
\caption{Peak positions (a) and full widths at half maximum (b) for $^{19}$F NMR spectra of BF$_4^-$ anions in the [BMI][BF4]/ACN ``same-lattice" systems.}
\label{fwhm-2-site-realistic}
\end{figure}

For the ``same-lattice" configuration, some NMR spectra show one peak while others show two (for particle sizes 14 and 15). The comparison between the 2-site and the realistic models for the peak position and full width at half maximum is shown in Figure~\ref{fwhm-2-site-realistic}. While the peak position varies smoothly in the 2-site model, it is not varying much in the realistic model up to the point where two peaks are observed. As for the ``same-ratio" configuration, the quantity of ions in the particle is smaller than in the bulk, explaining this trend. The evolution of the full width at half maximum is also markedly different for the two models.

The comparison between the realistic and 2-site models highlights the necessity of representing different quantities of ions in the particle and in the bulk, and the variety of chemical shifts, in order to predict NMR spectra comparable with experiments. 

\section{Conclusions}

In this work, we describe the implementation of a mesoscopic lattice model to study the effect of particle size on the NMR spectra of species adsorbed and diffusing in porous carbon materials. Distinct regions are defined to represent the bulk electrolyte and the electrolyte confined in the carbon pores. This approach allows one to account for the in-pore/ex-pore exchange in addition to the intra-pore exchange. Our investigations with a 2-site model underline that pre-existing analytical models for exchange between two chemically different environments are not sufficient to explain the spectral features observed experimentally for the adsorbed species. This is a direct consequence of the complex exchange processes present in the system. The comparison between the 2-site and realistic models highlights the importance of representing a range of magnetic environments and a range of chemical exchange rates to predict NMR spectra in a realistic fashion. This work opens the door to more precises NMR spectra prediction for specific systems, allowing for a more detailed interpretation of experimental results. The inclusion of several particles, with various sizes, is a possible perspective of this study.

\section*{Acknowledgements}

This project has received funding from the European Research Council (ERC) under the European Union's Horizon 2020 research and innovation program (grant agreement no. 714581). This work was granted access to the HPC resources of the CALMIP supercomputing centre under the allocations P21014, and of TGCC under the allocation A0090911061 made by GENCI. The authors acknowledge Alexander Forse, John Griffin, and Patrice Simon for fruitful discussions.

\section*{Author contributions}

Anagha Sasikumar: Methodology, Software, Validation, Investigation, Formal analysis, Writing – Original Draft, Writing – Review \& Editing. Céline Merlet: Methodology, Software, Validation, Writing – Original Draft, Writing – Review \& Editing, Supervision, Resources, Funding acquisition.



\section*{Research data}

The program used to do the lattice simulations is available, along with a manual, on Github (https://github.com/cmerlet/LPC3D). The data corresponding to the plots reported in this paper, as well as input files for the lattice simulations, are available in the Zenodo repository with the identifier 10.5281/zenodo.7895313.



\end{document}